\documentclass[fleqn,12pt,twoside]{article}
\usepackage{espcrc1}

\usepackage{graphicx}
\usepackage[figuresright]{rotating}

\newcommand{\nabsq}{\overrightarrow{\nabla}^{2}\!}

\newcommand{\beq}{\begin{equation}}
\newcommand{\eeq}{\end{equation}}
\newcommand{\bea}{\begin{eqnarray}}
\newcommand{\eea}{\end{eqnarray}} 

\newcommand{\AmS}{{\protect\the\textfont2
  A\kern-.1667em\lower.5ex\hbox{M}\kern-.125emS}}

\title{Effective Range Corrections in Few-Body Systems with large
Scattering Length}

\author{L. Platter\\[2mm]
{Department of Physics \& Astronomy, \\
        Ohio University, \\ 
        Athens, OH 45701,\\
        U.S.A.}}
       
\begin{document}

\maketitle

\begin{abstract}
The effective field theory with contact interactions alone is a
powerful tool to compute low-energy observables for three-body systems with
large scattering length. Recent calculations including effective range
corrections are discussed and results are presented.
\end{abstract}

\section{Introduction}
Effective field theories (EFT) exploit an existing separation of scales and
describe physical systems with a minimal set of degrees of freedom.
In the last years the effective field theory with contact
interactions alone (CEFT) has been applied very successfully
to various few-body systems with large scattering length \cite{Bedaque:1998kg,Braaten:2004rn,Platter2005}.
Furthermore, recent publications concentrated on the inclusion
of higher order corrections in calculations for three-body
observables \cite{Bedaque:2002yg,Griesshammer:2004pe}.
However, an obvious problem within this
framework, is the question on the power counting of many-body forces.
While a consistent power counting has been used in previous calculations
for the three-nucleon system with great success \cite{Bedaque:2002yg},
we have shown recently \cite{Platter:2006ev} that an energy-dependent
three-body force can be shifted down by one order
when using a subtraction formalism to compute observables. This allows
to analyze two-parameter correlation plots to one order higher than 
previously thought and improves the predictive power of the CEFT.

In the following two sections we will report on recent results obtained
within this framework. The last section summarizes our findings and gives an
outlook for further possible calculations.
\section{The Three-Boson System}
The CEFT is formulated in terms of the available degrees of freedom, {\it
  i.e.} heavy boson or fermion fields only and is valid if the underlying
potential is short-ranged and the involved momentum is smaller than the
inverse range of the potential. The most general Lagrangian describing
non-relativistic bosons interacting through contact interactions only
is given by
\bea
  {\cal L}  &=&
       \psi^\dagger \biggl[i\partial_t + \frac{\nabsq}{2M}\biggr]
                 \psi - \frac{C_0}{2}(\psi^\dagger \psi)^2
            - \frac{D_0}{6}(\psi^\dagger\psi)^3
                                           +  \ldots
\,,\label{lag}
\eea
where the ellipses denote interactions with more derivatives
and/or more fields.
With the corresponding power counting this EFT is an expansion in $R/a$,
where $R$ denotes the range of the underlying potential and $a$ the
two-body scattering length.
It is a particular feature of this theory that when applied to the the three-body
sector, a three-body force is needed already at leading order \cite{Bedaque:1998kg}.
However, instead of using an explicit three-body force one can use alternatively
a subtraction formalism which trades the three-body force in the
kernel of the three-body Faddeev equation for an explicit
three-body observables in the inhomogenity of the integral equation
\cite{Afnan:2003bs,Hammer:2000nf}.
It was shown recently that when this framework is used an energy
dependent three-body force needs not to be included at
next-to-next-to-leading order (NNLO) \cite{Platter:2006ev}.

This allows in particular to describe observables of $^4$He
few-body systems to very high precision.
Since the atom-atom scattering length is
$a_2\sim100$\AA\, while the typical van der
Waal's forces between the Helium atoms have a range $R\approx 10$~\AA,
the EFT expansion parameter is $R/a_2 \approx 0.1$ and
the EFT expansion should converge very quickly. In absence of any
experimental information about the binding energies of $^4$He trimers
we use theoretical few-body calculations using ``realistic'' atom-atom 
potentials \cite{RoY00,Ro00,MSSK01,Kolganova01}
to fix our three-body input and for comparison with our results.
Therefore, when using $a_{Ad}=1.205~\gamma^{-1}$ as obtained by Roudnev for the TTY potential
we obtain $B_3^{(0)}= 89.38~B_2$ for the ground state and $B_3^{(1)}=1.737~B_2$ for the
excited trimer state at NNLO. This is in very good agreement with
Roudnev's results for the ground and excited state,
$B_3^{(0)}=96.33~B_2$ and $B_3^{(1)}=1.738~B_2$, respectively.
\begin{figure}[t]
\centerline{\includegraphics*[width=12cm,angle=0]{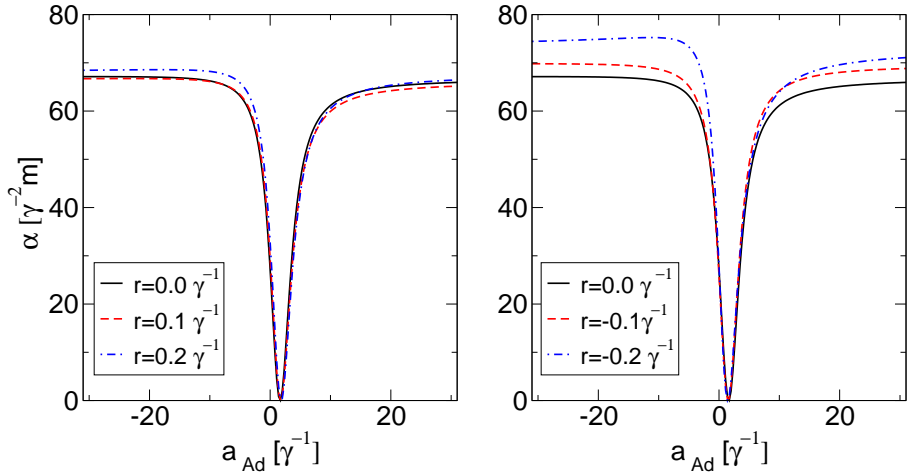}}
\caption{\label{fig:recombination} The left and right plot show the
  recombination coefficient $\alpha$ plotted against the atom-dimer
scattering at NLO. In the left plot results are plotted for
positive effective ranges $r=0.0,\ 0.1~\gamma^{-1}$ and $0.2~\gamma^{-1}$
(solid, dashed and dot-dashed line, respectively).
In the right plot results are plotted
for negative effective ranges $r=0.0,\ -0.1~\gamma^{-1}$ and
$-0.2~\gamma^{-1}$ (solid, dashed and dot-dashed line, respectively).}
\end{figure}
A further process one can consider within this framework is a common
reason for atom losses in experiments with ultracold alkali gases;
namely the recombination of three atoms into a shallow dimer and an atom.
At leading order (LO) the recombination rate has been calculated
by Braaten {\it et al.} \cite{Bedaque:2000ft}. 
In Fig.\ref{fig:recombination} we show our results for the correlation
between the atom-dimer recombination rate and the atom-dimer scattering length
at next-to-leading order (NLO) \cite{HLP}. In the left graph we display
results for different negative values of effective range and in the right graph
for different positive values. While all curves show a minimum in the
recombination coefficient at approximately $a_{Ad}\approx 1.7\gamma$, the
maximum value of $\alpha$ depends strongly on the sign and absolute
values of the effective range.
\section{The Three-Nucleon System} 
\begin{figure}[t]
\centerline{\includegraphics*[width=7cm,angle=0]{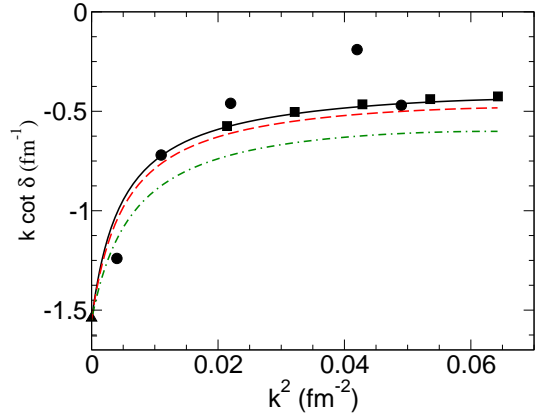}}
\caption{\label{fig:nd_kcotd}Phase shifts for neutron-deuteron scattering
below the deuteron breakup at LO (dashed-dotted line), NLO (dashed line)
and NNLO (solid line). The triangle is the result of the scattering
length measurement of \cite{Huffman:2005jx}. The circles are the results
of the van Oers-Seagrave phaseshift analysis \cite{seagrave}, and the squares denote
a phaseshift calculation using a realistic nucleon-nucleon
potential~\cite{Kievsky:1996ca}}
\end{figure}
The formalism introduced above can easily be applied to fermions and
extended to include spin and isospin degrees of freedom.
If this is done one is able to compute three-nucleon observables
in the triton channel up to NNLO with
five parameters \cite{Platter:2006ad}; the singlet and triplet
scattering lengths,
the corresponding effective ranges and one three-body parameter.
For the triton binding energy we obtain $B_t^{(\hbox{\tiny LO})}=8.08$~MeV,
$B_t^{(\hbox{\tiny NLO})}=8.19$~MeV at LO and NLO, respectively and
$B_t^{(\hbox{\tiny NNLO})}=8.54$~MeV at NNLO if we match to
the neutron-deuteron scattering length $a_{nd}=0.65$~fm \cite{Huffman:2005jx}.
We have also calculated the corresponding phaseshifts and our results are
shown in Fig.\ref{fig:nd_kcotd}.
For comparison we show in the same figure the results of a forty year old
phaseshift analysis \cite {seagrave} and a theoretical calculation using a realistic
nucleon-nucleon potential \cite{Kievsky:1996ca}. At higher order our
results seem to describe the experimental data better but considering
the age of the analysis and the fact that no errors are given for these data,
the theoretical calculation  by Kievsky {\it et al.} should be
considered as the true benchmark test for our calculation.   
At NLO our results already lie significantly closer to this calculation
and nearly perfect agreement is achieved at NNLO.
It should be noted that our results at LO and NLO order agree
with previous EFT calculations results given in
\cite{Hammer:2001gh,Bedaque:2002yg,Afnan:2003bs}. We also achieve qualitative
agreement at NNLO with Ref.\cite{Bedaque:2002yg}, however, without
employing an additional three-body counterterm.

\section{Summary}
The CEFT is designed for the calculation of low-energy observables
for systems with short-range interactions. When applied to systems
with large scattering length, the EFT expansion parameter is $R/a$.
It is universal in the sense that it can be applied to any
short-ranged interaction at low enough energies and therefore, it is
particular well suited to compute observables in atomic and
nuclear systems which have a large two-body scattering length.
Here, we have dicussed results for low-energy three-body observables
with effective range correction included up to NNLO obtained
without including a second three-body datum at this order as
dictated by a 
renormalization group analysis given in \cite{Platter:2006ev}.
This increases the
predictive power at this order as observables are described in terms
of one renormalization parameter less as has been done in previous
calculations in the CEFT. 
Our results are showing the expected convergence behavior and agree 
very well with previous theoretical calculations and experiment.
It should be noted that this
doesn't  indicate that results previously obtained at NNLO are wrong
but rather that they can be obtained with less input information.
Further effort should be devoted on the calculation of electromagnetic
properties with the CEFT \cite{Platter:2005sj}.

\end{document}